# Football tracking networks: Beyond event-based connectivity

J.M. Buldú[1,2], D. Garrido[1,2], D.R. Antequera[1,2], J. Busquets[3], E. Estrada[4,5], R. Resta[6] and R. López del Campo[6]

[1] Laboratory of Biological Networks, Center for Biomedical Technology, UPM, Madrid, Spain
[2] Complex Systems Group & GISC, Universidad Rey Juan Carlos, Móstoles, Spain
[3] ESADE Business School, Barcelona, Spain
[4] Instituto Universitario de Matemáticas y Aplicaciones, Universidad de Zaragoza, Spain
[5] ARAID Foundation, Government of Aragón, Zaragoza, Spain
[6] Mediacoach - LaLiga, Madrid, Spain

*ABSTRACT*

We propose using Network Science as a complementary tool to analyze player and team behavior during a football match. Specifically, we introduce four kinds of networks based on different ways of interaction between players. Our approach's main novelty is to use tracking datasets to create *football tracking networks*, instead of constructing and analyzing the traditional networks based on events. In this way, we are able to capture player interactions that go beyond passes and introduce the concepts of (a) Ball Flow Networks, (b) Marking Networks, (c) Signed Proximity Networks and (d) Functional Coordination Networks. After defining the methodology for creating each kind of network, we show some examples using tracking datasets from four different matches of *LaLiga Santander*. Finally, we discuss some of the applications, limitations, and further improvements of football tracking networks.

*INTRODUCTION*

More than three decades ago, Peter Gould and Anthony Gatrell introduced a groundbreaking concept in football teams' analysis: Transforming team dynamics into a network [Gould & Gatrell, 1979]. They recorded the final of the 1977 FA Cup between Manchester and Liverpool and extracted, manually, all passes between players of both teams. Next, they divided passes into two groups, one per team, and considered each player to be a node of a network. Nodes were connected between them through links, whose weight consisted of the number of passes made between each pair of players. In this way, Gould & Gatrell created the first (ever) football passing networks, one for Manchester United and one for Liverpool. The next step was to analyze them. They chose a methodology known as q-analysis [Atkins 1974; Gould 1980], which consists of characterizing the topology of networks by analyzing (1) the set of elements conforming a network and (2) their hierarchical organization. Despite its novelty and possible applications in football analysis, the paper of Gould & Gatrell did not have a very high impact in the community of sports sciences and never reached the pitch.

Nowadays, the situation has changed completely. The access to detailed datasets containing all actions occurring during a match, even the position of players and the ball at any moment, has promoted a diversity of new methodologies to describe and understand what happens on the pitch [Gudmundsson and Horton, 2017]. On the one hand, the analysis of event datasets has been extended in the communities of sports scientists, club analysts and even journalists. This kind of datasets, provided by a diversity of companies such as *StatsPerform*, *StatsBomb*, *Wyscout*, *InStat* or *Mediacoach*, to name a few, consist of every single action occurred during a match with (a) the spatial and time coordinates, (b) the result of the action and (c) the players involved. Importantly, event datasets have allowed developing new key performance indicators (KPI) that go beyond classical statistics such as the percentages of possession, completed passes, or shot success. For example, it is possible to calculate the expected goals (xG) of a team during a match by taking into account the coordinates of each shot together with the angles to the opponent's goal and the position of the defenders [Green, 2012; Lucey et al., 2015; Link et al., 2016]. Based on statistics from different competitions, xG accounts for the probability of scoring each shot, no matter what the final result is. In the same vein, other parameters such as the Post-Shot expected Goals (PSxG) [Goodman, 2018], the eXpected Thread (xT) [Singh, 2018] or the Goals Saved Above Average (GSAA) [Knutson, 2018], quantify the value of player actions based on event datasets.

All these "advanced" metrics refer to actions performed by a player, however, what happened to the pioneering idea of Gould & Gatrell of using networks to analyze football teams? The answer is that the access to event datasets has also fostered the amount and depth of network-based analysis of football. During the last years, passing networks have been widely used in the scientific community, clubs, and media. The paper of Duch et al. [Duch et al., 2010] quantifying the performance of players using network-based key performance indicators (*KPIs*) is considered to be the renaissance of the application of Network Science [Barabási, 2016; Newman, 2018] to the analysis of player and team behavior on the pitch, although some previous results were already preparing the scientific community to the use of networks in football [Onody & De Catro, 2004; Bundio & Conde 2009].

Since then, a diversity of network models has been proposed for capturing the complex patterns of connections between players of a team. For example, in *player passing networks* [Passos et al., 2011; Grund, 2012; Buldú et al. 2018], the two essential elements of a network (nodes and the links between them) are obtained precisely in the same way as Gould and Gatrell proposed in their seminal paper. In the example of Figure 1(A), we plot players (nodes) at the average position of all their passes, and we connected them through links whose width is proportional to the number of passes between them. A single image gives information about how all players passed the ball between them, but also how the team was placed on the pitch, its closeness to the opponent's goal, or the preference of attacking using the left or the right wings. Interestingly, the information contained in the player passing network has also been used for defining player *KPIs* [Duch et al., 2010; López-Peña et al., 2012; Cotta et al., 2013] or even to find players with similar features [López-Peña et al, 2015].

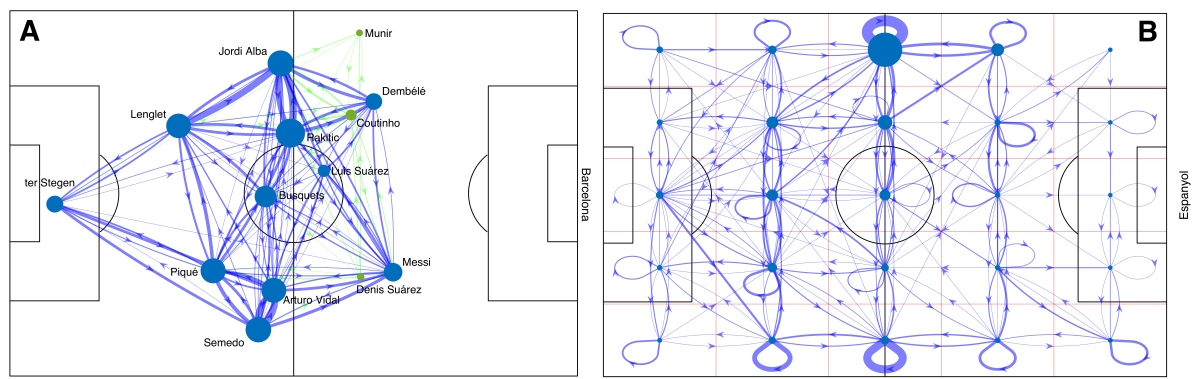

**Figure 1.-** Construction of event-based passing networks. In (A), we show an example of the player passing network obtained during the match between F.C. Barcelona (away team) and R.C.D. Espanyol (home team) during the 2018/2019 season. Players, who are the network's nodes, are placed at the average position of all their completed passes. Links account for the number of passes between pairs of players. Substitutes are highlighted in green. In (B), we plot the pitch passing network of the same match. In this case, nodes are divisions of the pitch and links account for the number of passes between them.

It is also possible to define other kinds of passing networks. In *pitch passing networks* [Cintia et al., 2015; Herrera-Diestra et al., 2020], nodes consist of partitions of the pitch that "pass" the ball between them without paying attention to the player making the pass (see Figure 1(B)). These networks maintain the spatial information about the starting and ending points of all passes, which was lost in player passing networks. By combining player and pitch passing networks we have a more complete description of how passes are related to the style of playing of football teams. There are more sophisticated ways of defining networks, such as pitch-player passing networks, where nodes are a combination of players and their positions on the pitch [Narizuka et al., 2014; Cotta et al., 2013] or hypergraphs, where nodes consist of a combination of players in a given context (e.g., two forwards facing one defender) [Ramos et al., 2017]. However, in all these examples, networks are obtained from event datasets, raising a natural question: Would it be possible to obtain networks from tracking datasets?

Despite the access to tracking datasets is still elusive to the majority of the scientific community, the last years have witnessed a series of new tacking-based metrics that suggest a revolution in the analysis of football in the years to come [Gudmundsson, J., & Wolle, 2014; Link et al., 2016; Marcelino et al, 2020, Carrilho et al, 2020]. Between all these metrics, the *pitch-control* is probably the one that has captured the most attention probably due to its simplicity and, at the same time, its ability to include information about how players use the space on the pitch, leading to the evaluation of actions beyond what is happening just around the ball [Taki & Hasegawa, 2000; Spearman et al., 2017; Fernández & Bornn, 2018]. However, the analysis of tracking datasets under the scope of Network Science is still a pristine field with more questions to make than answers given.

In this paper, we propose different alternatives for constructing networks of players based on tracking datasets. The fact that tracking goes beyond actions around the ball permits us to define new ways of interaction between players or areas of the pitch. These interactions will constitute the basis for creating new links between nodes, obtaining new information that will go beyond passing interactions. In the following, we will first describe the way tracking data is obtained. Next, we will introduce four new ways of obtaining different *tracking networks*. Finally, we will discuss what type of information can be extracted from each kind of network and suggest ways of developing and improving the construction of these networks.

## MATERIALS AND METHODS

The datasets we used to illustrate the construction of tracking networks have been supplied by *LaLiga* software *Mediacoach®* [Mediacoach 2020]. A multi-camera tracking system recorded each player's position on the pitch at 25 frames per second by using a stereo multi-camera system composed of two units placed at either side of the midfield line (Tracab Optical Tracking System) [Tracab 2020]. Each multi-camera unit contained three cameras with a resolution of 1920x1080 pixels that were synchronized to provide a panoramic picture and created the stereoscopic view for triangulating the players and the ball. An experienced operator corrected the position of players in the case of a temporal loss of any location. Importantly, datasets obtained by the *Mediacoach®* system have been previously validated with GPS [Garcia-Unanue et al. 2019; Pons et al., 2019].

Finally, the event datasets used to construct passing networks of Fig. 1 were supplied by Opta [StatsPerform 2020].

## RESULTS

### Ball Flow Networks

The movement of the ball along the pitch is crucial to evaluate the style of playing of a team. We propose using *Ball Flow Networks* (*BFN*) to understand how teams advance towards the opponent goal and to detect those regions of the pitch that are more relevant during the offensive phase. These networks contain the information about how the ball moves across the pitch when possessed by a team, without paying attention to the player who is conducting or passing the ball. The procedure to obtain a *BFN* from tracking datasets beings with the partition of the pitch into a *NxM* grid, similar to what is done to obtain pitch passing networks, as the one shown in Fig. 1(B). Each partition is a node of the network. Next, we count how many times the ball traverses the frontier between any pair of partitions and we use this information as the weight of the links between two nodes. Note that this procedure leads to one *BFN* for each team. In Fig. 2, we show the *BFN* of F.C. Barcelona during its match against R.C.D. Espanyol (season 2018/2019). To include more information, we plot the size of the nodes proportional to the time that the ball has been retained by a team inside each partition. Thanks to the arrows (i.e., the links between two nodes), we can see the preferred ways of moving the ball towards the opponent's goal. In this case, we observe how F.C. Barcelona moved the ball from its defensive lines departing from the center of its goal (which indicates a preference of play with the goalkeeper) and using the right side of the pitch a bit more than the left one. Once at the midfield, there is a preference for playing on the left wing of the pitch, as indicated by larger size nodes. Finally, it is worth paying attention to how the ball enters the opponent's area. The ball goes through the opponent's area through the right side a bit more frequently than the left one. However, both sides are less used than the front part, which contains the highest number of incursions.

Importantly, this kind of network can be adapted to the particular interests of scientists or clubs. For example, node size could be proportional to other variables, such as (a) the percentage of plays that cross a specific partition and finish into a shot or a goal or (b) the number of passes made from each region.

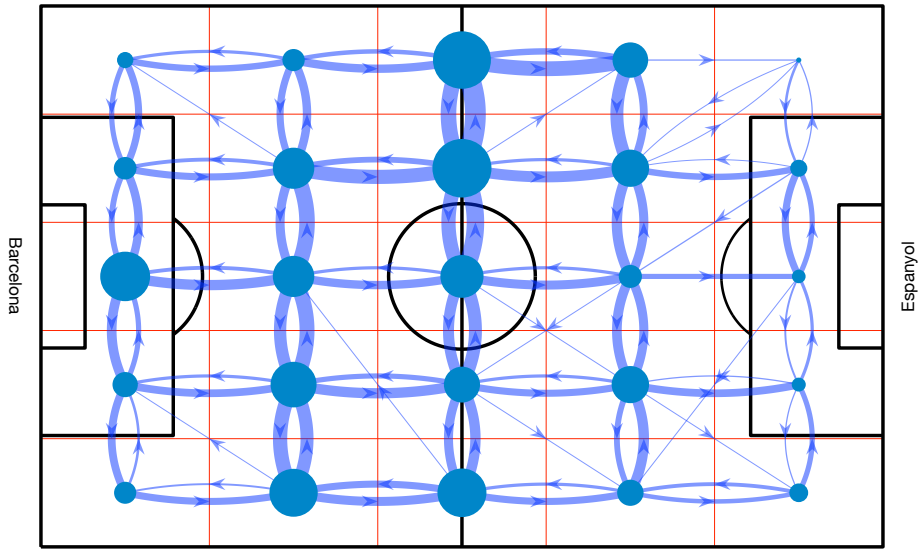

**Figure 2.-** Construction of a *Ball Flow Network* (*BFN*). In this example, we show the movement of the ball across the pitch during the match between R.C.D. Espanyol (home team) and F.C. Barcelona (away team), season 2018/2019. Only the possession of F.C. Barcelona is plotted (attacking from left to right). To construct the network, we divide the pitch into *NxM* partitions, in this case *N=M=5*. Each partition is a node of the network. Next, we count the number of times the ball goes through one partition to another and use this value as the weight of the links. The thickness of the links is proportional to the number of times the ball crossed the frontier between two partitions. The size of the nodes (circles) indicates the time the ball has been inside each partition during the attacking phase of F.C. Barcelona.

## *Signed Proximity Networks*

Understanding of the position of players referred to their partners and opponents is crucial to describe how teams organize both during the defensive and offensive phases. *Signed Proximity Networks* (*SPN*) aim to capture these spatial dependencies by constructing networks based on the distances between players. In this case, the nodes are the players themselves; however, we obtain a unique network for all players instead of one network for each team, as it was the case of (event-based) player passing networks. Furthermore, signed networks are time evolving and, therefore, they must be tracked along the different moments of the match. Let us explain the procedure about how to construct them. First, for every frame obtained during the match, we connect any two players with a link if they are below a threshold distance *R*, whose value can be modified from zero to any given value. Importantly, links have a positive value (*+1*) if they connect players of the same team and a negative one (*-1*) if the connected players belong to different teams. As we can see in the example of Fig. 3(A-B-C), for a given frame, different values of R lead to networks containing a different number of links and, therefore, different properties. In this way, a set *{r}* of signed networks can be assigned to a unique frame. When this procedure is repeated for all the *{t}* frames of the match, we obtain *{r}x{t}* signed networks, whose spatial properties depend on *{r}* and the temporal ones on *{t}*.

Once *SPN* are obtained, their analysis can offer different insights about the players' position on the pitch and their interplay with partners and opponents. For example, in Fig. 3 we have analyzed the formation of triangles below the threshold distance R during the match between Atlético de Madrid and Real Valladolid (season 2018/2019). As we can see, only 4 types of triangles are possible. Two of them contain only positive links, and are composed of either players of Atlético de Madrid (orange nodes) or players of Real Valladolid (blue nodes). The other two types of triangles contain mixed signs with one of the two teams having a higher number of players than the other. Figure 3(D) shows the average number of the 4 types of triangles obtained during the whole match. We can observe how Atlético players form more triangles in configurations where they have most players (either all positive or mixed). Interestingly, it is around R~7 meters where the difference between both teams is the highest, i.e., the distance where triangles of Atlético de Madrid dominate the most.

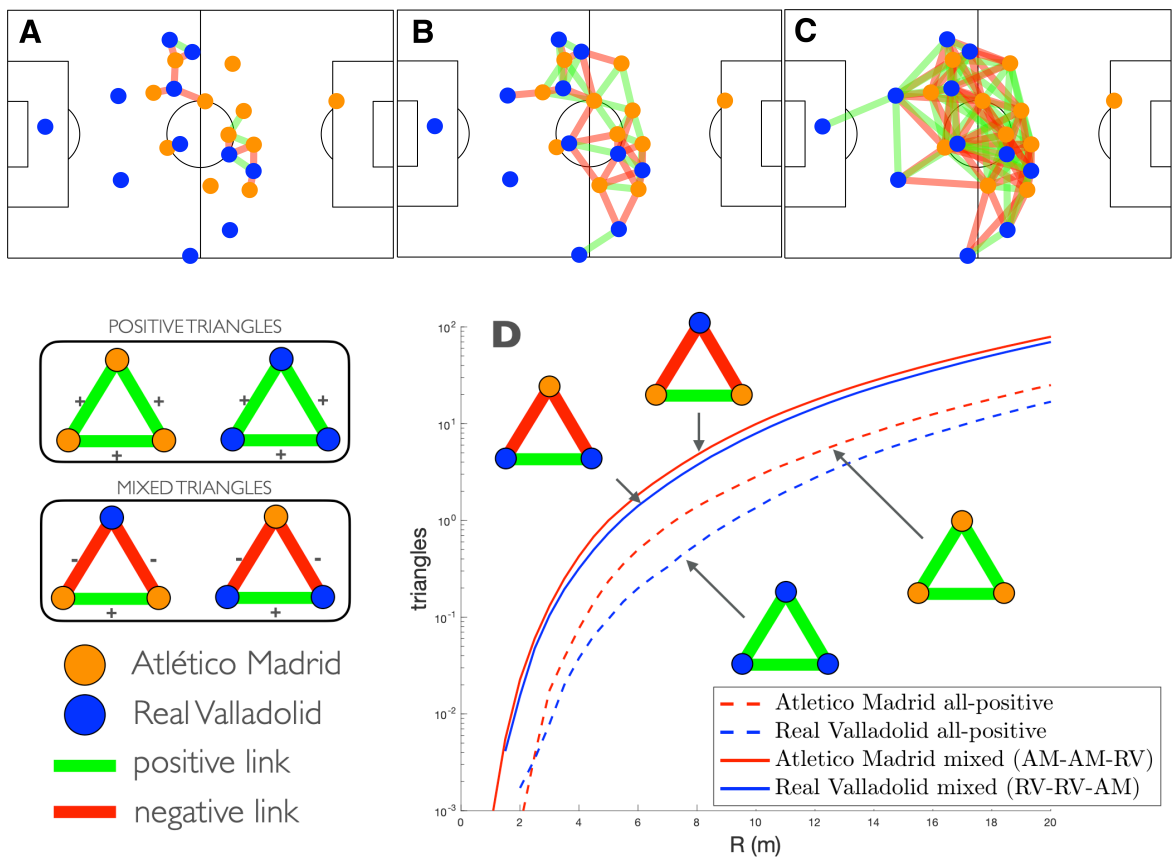

**Figure 3.-** Construction of a *Signed Proximity Network* (*SPN*). In the upper plots, an example of modifying the threshold distance *R* considered to create links between players: *R= 5* meters in (A), *R=10* meters in (B) and *R=15* meters in (C). The example corresponds to the match between Real Valladolid (RV, home team) and Atlético de Madrid (AM, away team) during season 2018/2019. Positive links are drawn in green and negative links in red. (Below) Example of the 4 kinds of triangles that can be constructed around the ball: two of them have only positive sings (positive triangles) and the other two have mixed signs (mixed triangles). In (D), average number of triangles as a function of the threshold distance *R* (in meters), for the whole match. Note that players of Atlético de Madrid form triangles with a higher probability than players of Real Valladolid at any threshold distance.

## *Marking Networks*

We can define another kind of tracking network by paying attention to a more specific interplay of the distances between players. As recently proposed by R. Tavares [Tavares 2020], we can focus on the time a defender devotes to mark a specific player by approaching him closer than a threshold distance *R* and construct a *Marking Network* (*MN*). For example, we can set *R=1.5* meters and find out the time accumulated by each player during the defensive phase with an opponent closer than *R*. This procedure leads to a bipartite network, i.e. a network composed of two different sets of nodes, each one being the players of a given team.

In bipartite networks, nodes of one set can only have links to nodes belonging to the other set. In other words, we only have connections between players of one team to players of the other team. The weight of these links is the time accumulated by a defender and his opponent below the threshold distance R. Figure 4(A) shows the connectivity matrix of the match between Real Madrid and Leganés (season 2018/2019). Precisely, this matrix corresponds to the *MN* of Leganés, which is different from the one of Real Madrid, the latter obtained by computing the accumulated times during the defensive phase of Real Madrid. In this particular example, the link with the highest weight corresponds to the mark made by Braithwaite (Leganés) to Casemiro (Real Madrid). Note that Braithwaite is a forward player and not a defender, however, his intense pressure over Casemiro during the match placed him in the first position surpassing the defenders (who follow him in the ranking). Figure 4(B) shows the corresponding marking network, where the nodes' position corresponds to the average position of the time accumulated during the marking, and the width of the links is proportional to the accumulated time between pairs of players. In this example, we removed those links with an accumulated time lower $t_{min}=7$ seconds (to highlight the most important connections). We can observe that Benzema was the most marked player, a task that was mainly divided between the defenders Bustinza, Omeruo and Siovas.

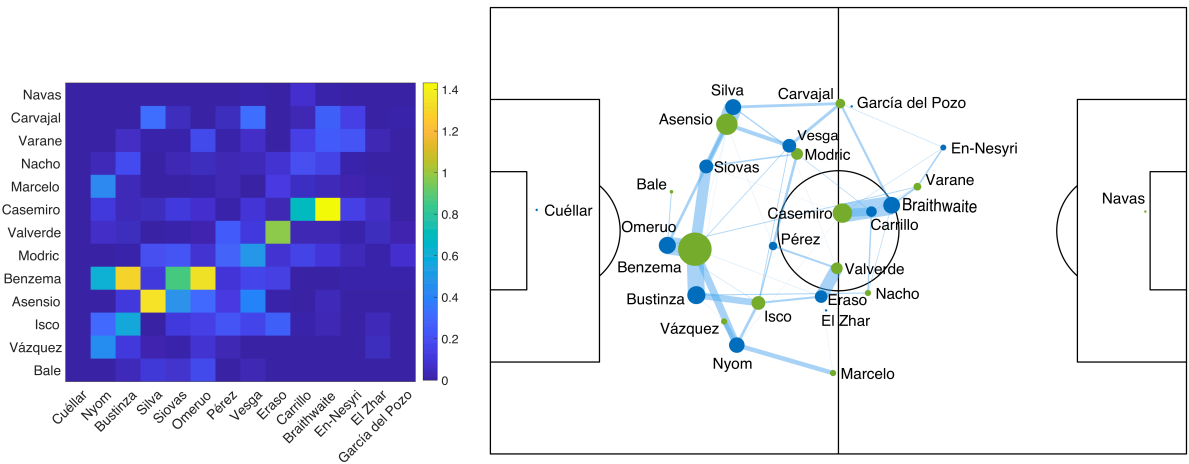

**Figure 4.-** Construction of *Marking Networks* (*MN*). In the left plot, connectivity matrix of the marking network corresponding to the defensive phase of Leganés (home team) during its match against Real Madrid (away team), season 2018/2019. Each element of the matrix corresponds to the time (in minutes) accumulated by a defender at a distance below *R=1.5 m* of any of his opponents. In the right plot, the corresponding marking network of Leganés. Nodes (players) are plot at the average position during the defensive phase and the width of the links is proportional to the accumulated time between pairs of players. Only links with an accumulated time higher than $t_{min}$=7 seconds have been plot.

*Functional Coordination Networks*

"*All models are wrong, but some are useful*", this quotation by George Box summarizes the purpose of a mathematical model: to be useful despite its limitations. With this regard, our last proposal is probably the one with more simplifications, requiring further development.

However, we are confident of its utility. Now, we are interested in capturing the interplay between players of the same team. We would particularly like to know whether two players are coordinated in their motions and construct what we call *Functional Coordination Networks (FCN)*. Functional networks are commonly used in neuroscience and consist of networks whose nodes are different brain regions connected between them if there is any kind of coordination between their dynamics [Bullmore & Sporns 2009; Papo et al. 2014]. Translating this concept to tracking networks, we require knowing when two players move in a coordinated way and quantifying this coordination. However, this is an unsolved problem, since it is highly complex to know whether two players are coordinating their motions in purpose.

Assuming this limitation, we make the first attempt to construct a functional coordination network in football by paying attention to the alignment of the velocity vector of players of the same team. Specifically, we compute the velocity of each player at each frame. Since velocity is a vector that is tangent to the movement of a player (i.e., it indicates the direction), we can identify the time two players of the same team are moving towards the same direction, assuming that they are coordinating their motions. This is a very rough approximation since, as in the brain, there are many different ways of coordination. For example, when a lateral defender moves forward to incorporate to a counter-attack, a midfielder may move backwards to cover the space left by the defender. Our model would not capture this kind of actions. However, we believe that we still obtain very inserting information measuring the alignment of velocities. In Fig. 5(A)-(B), we show the matrices of the *FCN* of Real Betis (away team) during its match against R.C.D Espanyol (home team), at the season 2018/2019. To compute the elements of the matrices we first calculate the velocities of all players $v_i(t)$ for every frame of the match, with $i=1,...N_a+N_b$, being $N_a$ and $N_b$ the number of players of Betis and Espanyol, respectively. For every frame, we compute the angle $\theta_{ij}$ of the velocities of each pair of players $i$ and $j$. When the angle is below a certain threshold (in this case $\theta_{ij} < \theta_{thresholf}=10°$) we consider that the velocities of player $i$ and $j$ were aligned during this frame. Finally, we compute the percentage (between 0 and 1) of time accumulated by every pair of players during the whole match. Figs. 5(A)-(B) show the matrices containing the values of the players of Betis obtained during the defensive (A) and offensive (B) phases. We can observe how values of coordination during the defensive phase (greenish) are, in general, higher than those reported during the attack (bluish). Specifically the average values are $|C_{deffensive}|=0.176$ and $|C_{offensive}|=0.152$. This is somehow expected, since defence is based on maintaining order according to a particular structure, which implies higher coordination between players of the same team. On the contrary, during the attack, players are prone to move more randomly to create disorder in the defensive mesh of the opponent.

Figures 5(C-D) show the plots of the *FCNs* corresponding to the connectivity matrices of Figs. 5(A-B). Betis players are located at their average positions during the defensive (C) and offensive (D) phases. In turn, players' size is proportional to the time that they have been coordinated with any of their partners. The width of the links is proportional to the coordination time between any pair of players. To ease the figure's interpretation, we only plot links with a coordination time higher than the average. We can observe how during the defensive phase, midfielders are the players that coordinate the most, followed by defenders. Interestingly, during the offensive phase, all Betis forwards reduce their coordination with their teammates, revealing movements in different directions.

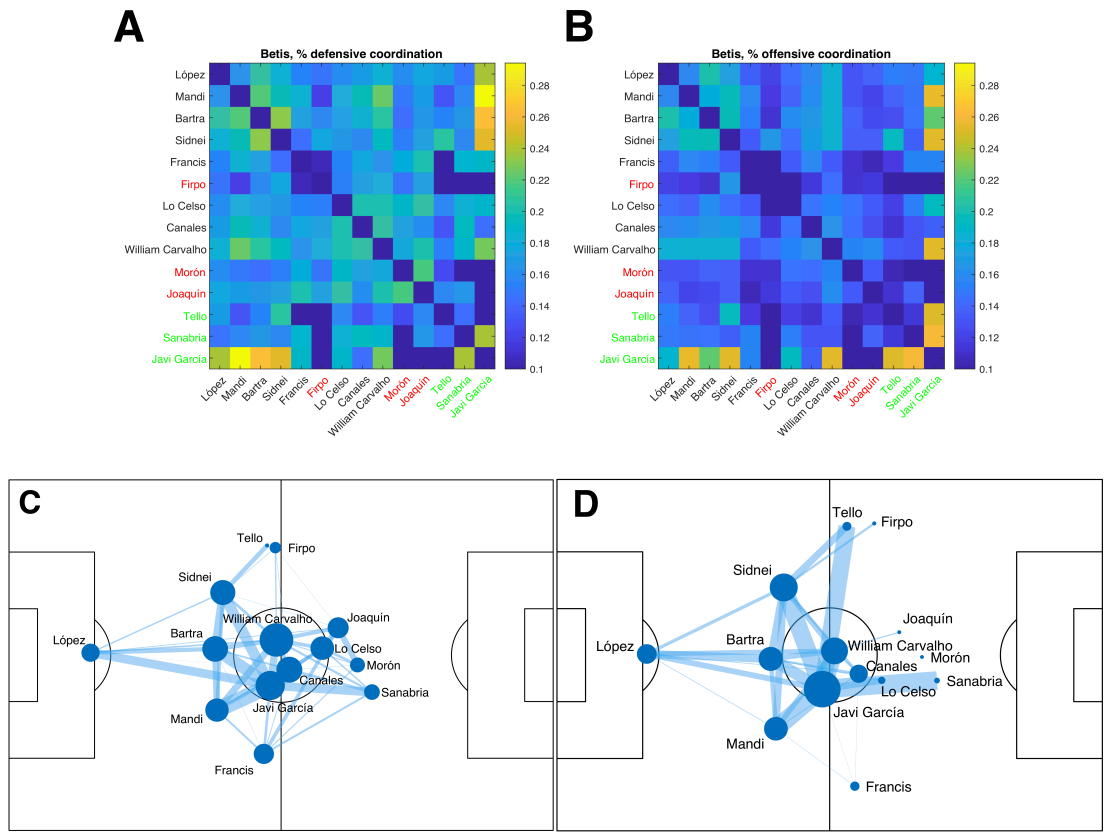

**Figure 5.-** Construction of *Functional Coordination Networks* (*FCN*). In the upper plots, coordination matrices of players of Real Betis (away) during its match against R.C.D. Espanyol (home team), season 2018/2019. In (A), the defensive phase is evaluated, while (B) contains the results of the offensive phase. In the bottom plots (C-D), we show the functional coordination networks of the matrices shown in (A) and (B). Nodes (all players who played the match) are located at their average position during their defensive (C) and offensive (D) phases. The size of the nodes is proportional to the coordination time accumulated by each player. The width of the links is proportional to the percentage of time that the players' velocities were aligned.

## *CONCLUSIONS*

We are witnessing a new era in the analysis of football datasets, where new technologies allow collecting information about players that had been inaccessible until the last years. However, it is the quality of the data and also the way it is being analysed what is making the difference with previous decades. Machine learning, complexity sciences or nonlinear dynamics are being introduced as tools for analysing football datasets, giving completely new perspectives about player and team performance.

In the current paper, we proposed to take the concept of football networks one step beyond, creating what we call *Tracking Networks*. We defined four new kinds of networks by using tracking datasets to extract the interactions between the nodes of the network, instead of events. In *Ball Flow Networks*, we tracked the ball's position and used it as a proxy to create connections between different areas of the pitch. This kind of tracking network allows understanding how a given team moves the ball during the attacking phases, highlighting those areas of the pitch that play a crucial role in the conduction towards the opponent's goal. They share some similarities with pitch passing networks, since the nodes are partitions of the pitch in both cases. However, ball flow networks contain not only passes but also the conduction of the ball by players. Furthermore, they show what are the regions of the pitch that are occupied the most during the attacking phase and how the ball arrives at (and leaves) them.

In *Signed Proximity Networks*, it is the distance between players that creates connections, leading to a player network that informs about the closeness of players with their partners and opponents. This kind of networks allows detecting local superiorities, indicating those teams whose players are closer between them and/or whose density of partners is higher than the density of rivals. Despite the information being based on local features (it only considers the distance to the closest partners), it also reveals global properties of a team. For example, when a given team has more positive links than its opponent, it indicates that its players are closer, leading to local superiorities through the formation of triangles of players. Furthermore, signed proximity networks could be used to track changes in the defensive and offensive phases, since the number of positive and negative links is highly dependent on possession.

The distance between players was also the feature to create connections between players in Marking Networks. In this case, we were concerned about how a player marks another (rival) player, and what is the time accumulated during the marking. Setting a threshold distance for detecting the proximity of the defender to the attacker, we construct bipartite networks that show how a team distributes the marking of the opponent players. When plotted, these networks allow seeing (a) the average position of markings, (b) the pairings that are more frequent or (c) how two or more players shared the marking of a given opponent.

Finally, we propose the use of *Functional Coordination Networks* as a preliminary way of analyzing how players of a given team coordinate their motions. Despite the way two players coordinate is highly complex, we proposed to analyze the amount of time that their velocities were aligned during a match. In this way, we obtained the functional coordination networks of the defensive and offensive phases. We used an example to show that the alignment of the velocities is higher at the defensive phase compared the offensive one. Furthermore, we observed that forwards have less coordination than the rest of their partners, especially during the offensive phase. Functional Coordination Networks still require including other types of coordination beyond the alignment of velocities, suggesting further improvements in the near future. For example, the use of the relative phase has been recently proposed to capture the synergy of players of a given team [Carrilho et al., 2020]. In this proposal, the each player's phase was obtained by measuring the angle of each player with the ball and the opponent's goal. Using the same approach, a phase synchronization index could be used to weight the links of functional coordination networks. More complex approaches, also using the alignment of the players' velocities have been used as a fingerprint to identify players and even to relate its coordination with teammates and opponents to their value in the transfer market [Marcelino et al, 2020]. Definitely, further improvements in the way players coordinate will give more precision to the construction of functional coordination networks.

In view of all, we are confident that the proposed kinds of tracking networks, and others to come, will be of extreme interest for scientists and football analysts interested in the performance of players and teams by using information that goes beyond event datasets.

*ACKNOWLEDGMENTS*

We would like to thank F. Seirul.lo (F.C. Barcelona), M. Conde (SuperLiga Argentina), J. Lagos (@Vdot_Spain) and M.A. Gómez Ruano (Universidad Politécnica de Madrid) for fruitful conversations about networks and football.